# Time Resolution of the 4H-SiC PIN Detector

Tao Yang[1,2], Yuhang Tan[1,2], Qing Liu[3], Suyu Xiao[1,2,4], Kai Liu[1], Jianyong Zhang[1], Ryuta Kiuchi[1], Mei Zhao[1], Xiyuan Zhang[1], Congcong Wang[1], Boyue Wu[5], Jianing Lin[6], Weimin Song[6], Hai Lu[3]* and Xin Shi[1]*

[1]Institute of High Energy Physics, Chinese Academy of Sciences, Beijing, China, [2]School of Physical Sciences, University of Chinese Academy of Sciences, Beijing, China, [3]School of Electric Science and Engineering, Nanjing University, Nanjing, China, [4]Shandong Institute of Advanced Technology, Jinan, China, [5]School of Physical Science and Technology, Guangxi University, Guangxi, China, [6]College of Physics, Jilin University, Jilin, China

We address the determination of the time resolution for the 100 μm 4H-SiC PIN detectors fabricated by Nanjing University (NJU). The time response to $\beta$ particles from a $^{90}$Sr source is investigated for the detection of the minimum ionizing particles (MIPs). We study the influence of different reverse voltages, which correspond to different carrier velocities and device sizes, and how this correlates with the detector capacitance. We determine a time resolution (94 ± 1) *ps* for a 100 μm 4H-SiC PIN detector. A fast simulation software, termed RASER (RAdiation SEmiconductoR), is developed and validated by comparing the waveform obtained from simulated and measured data. The simulated time resolution is (73 ± 1) *ps* after considering the intrinsic leading contributions of the detector to time resolution.

***Correspondence:**
*Hai Lu*
*hailu@nju.edu.cn*
*Xin Shi*
*shixin@ihep.ac.cn*

## INTRODUCTION

In the recent years, much attention has been devoted to seek the appropriate semiconductor material to be used in future particle colliders and nuclear reactors operating in harsh radiation environment (i.e., $> 10^{17}\ n_{eq}/cm^2$) [1]. Silicon-based detectors have the support of a sophisticated production technology and present good quality, but the leakage current sharply increases and charge collection efficiency rapidly decreases when the irradiation fluence exceeds $10^{15}\ n_{eq}/cm^2$ [2]. To enhance performance and lifetime, most of the silicon-based detectors also need an expensive cooling system, which makes the overall detector system giant and expensive. Alternative diamond detectors have been investigated with high radiation hardness up to $3 \times 10^{15}\ particles/cm^2$ [3] and have been successfully used in the ATLAS experiment at the LHC [4]. However, they are also characterized by high cost and a difficult doping process in diamond, which limit their application. On the other hand, the 4H-SiC material, owing to its potential high radiation hardness, wide bandgap energy (3.27 eV), high atomic displacement energy (25 eV), and stability at high temperature, has great potential for application in extreme radiation environments.

Owing to the currently achieved high-quality 4H-SiC epitaxy wafer, a handful of studies about charge collection, leakage current, capacitance, and deep energy levels of 4H-SiC detectors have been carried out before and after irradiation [5, 6]. Due to its wide bandgap energy, which is also insensitive to visible light, SiC detectors are useful for X-ray and ultraviolet monitoring [7]. There have also been extensive studies about the SiC detector's application in neutron detection in fusion devices [6, 8]. Concerning applications in high-energy physics, the detector's response to the MIPs is more relevant, but most of the previous studies focused on energy resolution and charge collection efficiency using $\alpha$ particles, which are distinct features from the MIPs. To the best of our knowledge, charge collection features relevant to MIPs in SiC detectors have been analyzed [9], but no investigations about time resolution have been reported, except for the 4H-SiC Schottky barrier diode (SBD), where time resolution with $\alpha$ particles for deuterium–tritium (D-T) applications has been analyzed [10].

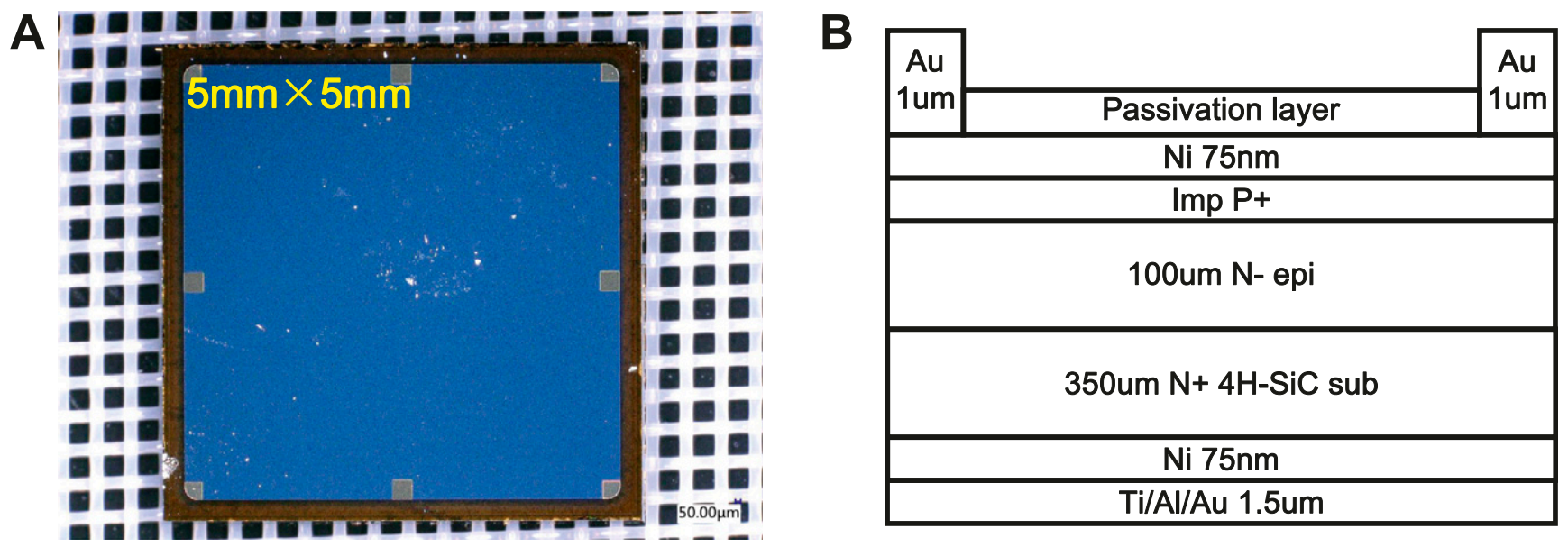

**FIGURE 1 |** 5 × 5 *mm* 4H-SiC PIN sample developed in NJU **(A)** Photography and **(B)** cross section. It has a 100 μm 4H-SiC active epitaxy layer and 350 μm substrate.

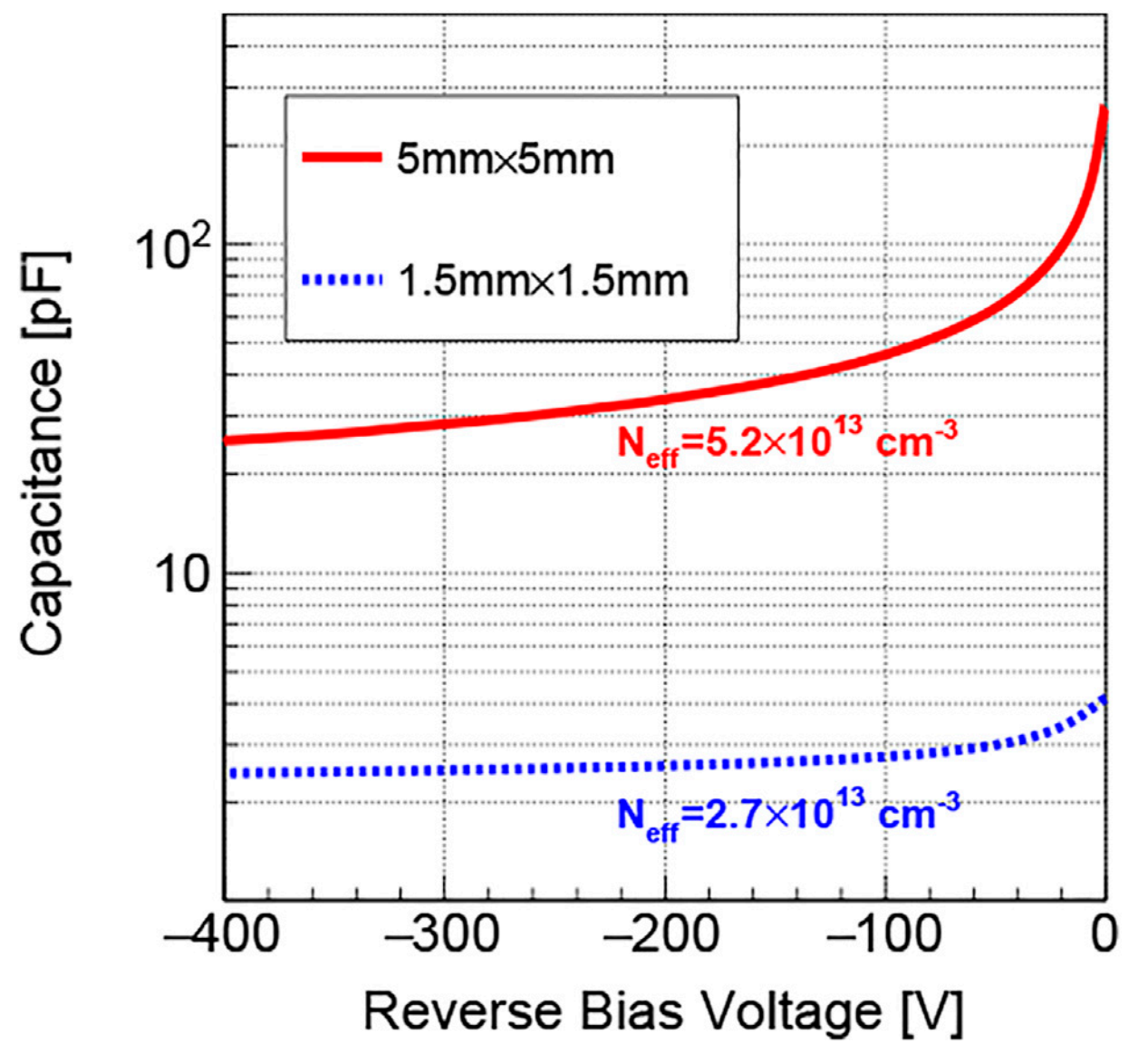

**FIGURE 2 |** Capacitance—voltage characteristics of two 4H-SiC PINs with different sizes. The capacitance is measured by a Keysight E4980A LCR Meter with a custom high bias voltage adapter (±400 V).

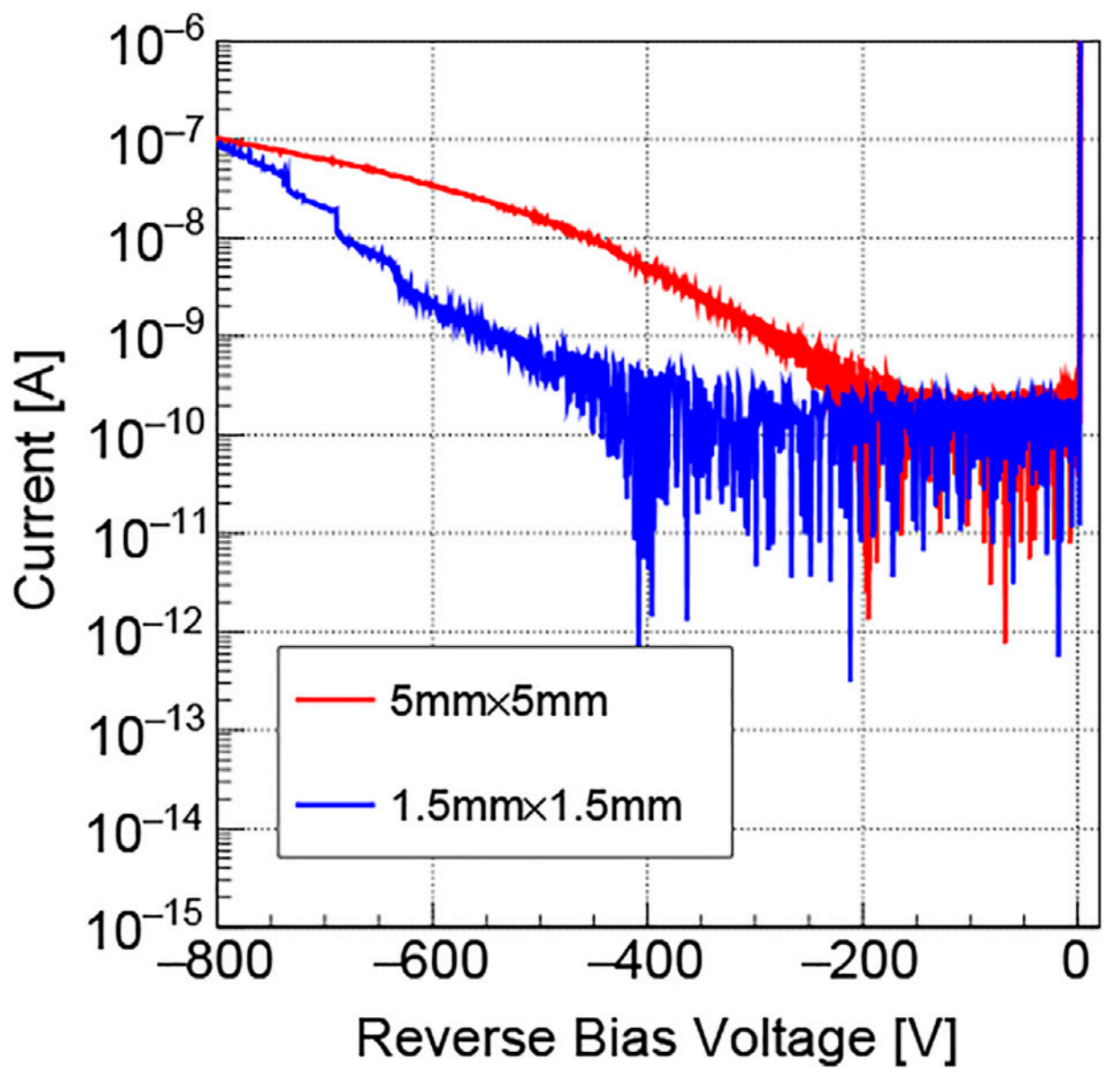

**FIGURE 3 |** Current—voltage characteristics of two 4H-SiC PINs with different sizes, as collected by a Keithley 2410 Source Meter.

In the recent years, ultrafast detectors have become a hot research topic. The main goal is to achieve a very high time and position resolution, simultaneously. A time resolution better than 20 *ps* has been achieved in silicon planar sensors with depletion thicknesses 133–285 $\mu m$ for multiple MIP signals [11], whereas 100 μm silicon pixel detectors with 800 $\mu m$ × 800 $\mu m$ size have achieved a time resolution of 106 *ps* [12]. Currently, 50 μm silicon detectors with internal gain, usually referred to as Low Gain Avalanche Detector (LGAD), are developed by various foundries and show a time resolution of at least 50 *ps* [13–18]. The 4H-SiC detectors also show fast time response, coming from the highly saturated carrier velocity, but no time performance study has been reported so far.

Motivated by the abovementioned arguments, we here investigate the time resolution of the 4H-SiC PIN device using a $^{90}$Sr source for applications in high-energy physics experiments.

A fast simulation environment to investigate time resolution is a beneficial tool to develop fast detectors and properly understand time response features. The present open-source software, for example, Weightfield2 [19] and KDetSim [20] are only available for silicon detectors. The corresponding simulation tool for silicon carbide detectors is lacking due to distinct material parameters. Therefore, we also developed a fast simulation software termed RASER [21] for applications with silicon carbide detectors, which was used in this study to reproduce measured data.

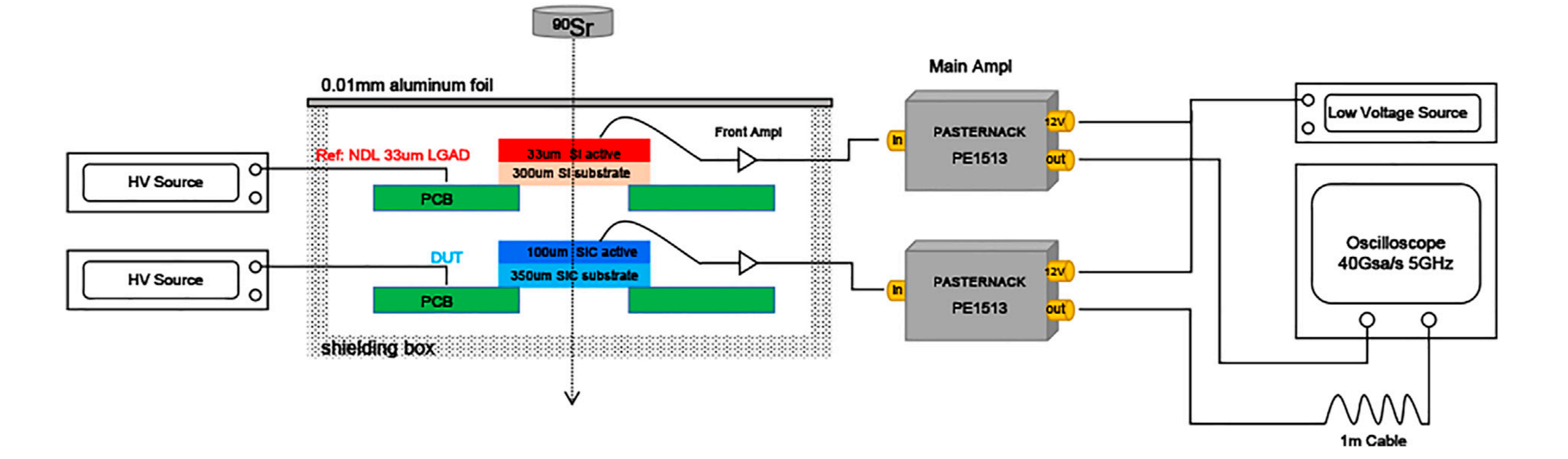

**FIGURE 4 |** Schematic diagram of the $\beta$ source test system to measure the time resolution. A 33 μm NDL Si-LGAD is used as a reference for timing. The front amplifier and main amplifier have 100 times to the input signal. Another 1 m cable connects with the output of the main amplifier to delay the DUT signal.

## THE DEVICE UNDER INVESTIGATION

The 4H-SiC PIN devices under investigation are fabricated by Nanjing University and come in two different sizes: 5 $mm \times 5$ $mm$ and 1.5 $mm \times 1.5$ $mm$. **Figure 1** shows the 5 $mm \times 5$ $mm$ sample which has two ohmic contacts on the top and bottom. The two devices both have a 100 μm high resistive active 4H-SiC epitaxy layer and 350 μm substrate, whereas the effective doping concentration is different: we have $N_{eff}$ (5 $mm \times 5$ $mm$) = 5.2 $\times 10^{13}$ $cm^{-3}$ and $N_{eff}$ (1.5 $mm \times 1.5$ $mm$) = 2.7 $\times 10^{13}$ $cm^{-3}$, respectively. These values may be extracted from the capacitance–voltage curve (see **Figure 2**) by

$$N_{eff} = \frac{2}{q\varepsilon A^2 d(1/C^2)/dV}, \tag{1}$$

where $q$ is the electron charge, $\varepsilon$ is the dielectric constant of 4H-SiC, and A is the area of the active region. Taking into account the doping levels and the dependence on device thickness $V_{dep} = \frac{q|N_{eff}|d^2}{2\varepsilon}$ with d = 100 $\mu m$, the full depleted voltages are given by $V_{dep}$ (5 $mm \times 5$ $mm$) = 484 $V$ and $V_{dep}$ (1.5 $mm \times$ 1.5 $mm$) = 248 $V$.

The current–voltage characteristics are shown in **Figure 3**. The typical unidirectional conduction characteristic of PINs is observed, and the breakdown voltage is larger than 800 V for both sizes. A higher leakage current may be collected by the 5 mm×5 mm size device due to its larger volume. The leakage current density is J < 100 $nA/cm^2$ for both the devices with a 500 V reverse voltage, where J (5 $mm \times 5$ $mm$) = 63.2 $nA/cm^2$ and J (1.5 $mm \times 1.5$ $mm$) = 34.6 $nA/cm^2$. The lower current density of the smaller device agrees with its lower effective doping concentration, as obtained from data in **Figure 2**.

## EXPERIMENTAL SETUP

### $\beta$ Source Test System

A schematic diagram of the experimental setup to determine the time resolution of 4H-SiC detectors is shown in **Figure 4**. We choose a 33 μm silicon LGAD as the reference timing device owing to its 34 ps time resolution at U = 200 $V$ and room temperature. The reference timing device is developed by the Institute of High Energy Physics (IHEP) of Chinese Academic Sciences and the Novel Device Laboratory (NDL) of Beijing Normal University [22–24]. The $^{90}$Sr source emits $\beta$ particles at 0.546 MeV from $^{90}$Sr and at 2.280 MeV from $^{90}$Y. Both are able to penetrate the LGAD device and deposit energy in the 4H-SiC detector. The front side readout boards are designed for LGAD devices by the University of California, Santa Cruz (UCSC). Each board has a 2 mm diameter hole in the middle. The overall front side electronics are transferred into a metal box to shield it from electromagnetic interference, with a 0.01 mm aluminum foil covering. Two 20 dB broadband amplifiers are placed before the oscilloscope to enhance the SNR. There is an additional 1 m cable on the DUT side to delay the signal (by ~5 ns) and enhance trigger efficiency. The sampling rate of the oscilloscope is 40 GSa/s, and each channel has 20 GSa/s.

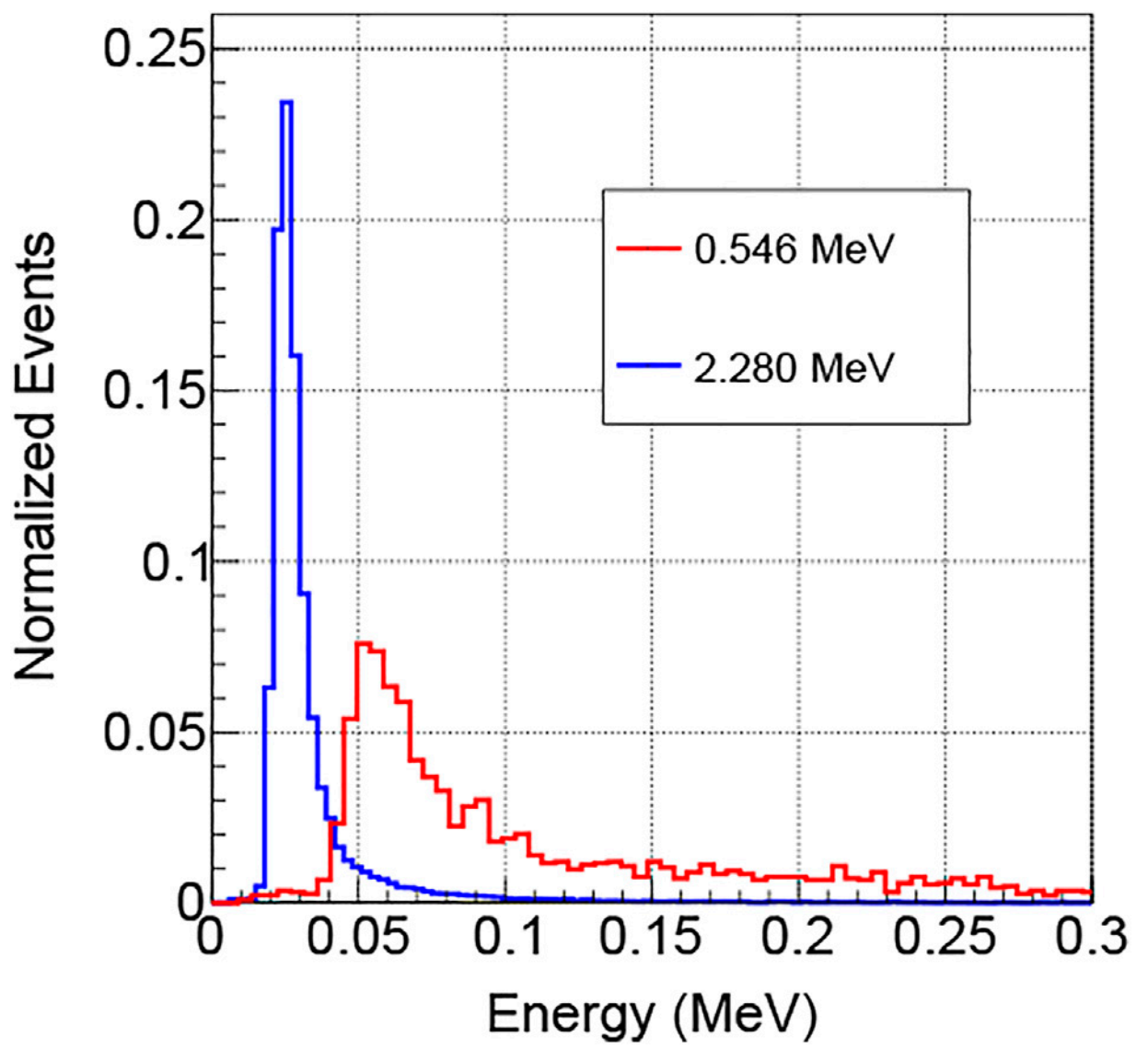

**FIGURE 5 |** Energy of $\beta$ particles from the $^{90}$Sr source deposited in the 100 μm 4H-SiC PIN active layer after they pass the aluminum foil and LGAD in GEANT4 simulation. The 0.546 MeV $\beta$ particles are from the $^{90}$Sr decay and the 2.28 MeV $\beta$ particles are from $^{90}$Y decay.

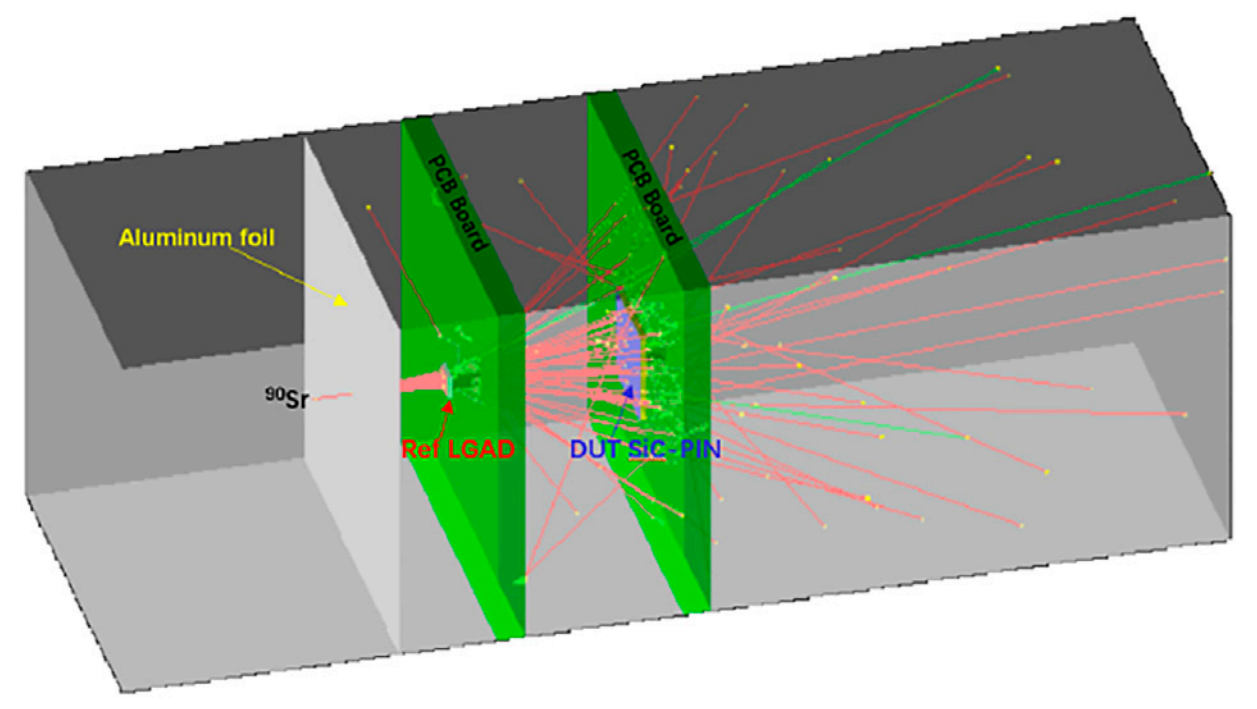

**FIGURE 6 |** Scattering of $\beta$ particles with 2.28 MeV after the aluminum foil and LGAD in GEANT4 simulations. The figure shows the tracks of 50 incident particles going through the entire $\beta$ source test system.

## Energy Response by GEANT4 Simulation

To analyze the energy loss of MIPs in 100 μm 4H-SiC active layer tallies, we use a simulation based on GEANT4, which allows one to describe the energy deposition. **Figure 5** describes the energy deposition of the $\beta$ particle in the 100 μm 4H-SiC active layer in the system (see **Figure 6**). The energy loss in the aluminum foil may be neglected. The scattering of $\beta$ particles in the aluminum foil and LGAD strongly decreases trigger efficiency for the 4H-SiC device. In turn, this explains the difference of trigger efficiency between the $5\,mm \times 5\,mm$ (4.7 events/min) and the $1.5\,mm \times 1.5\,mm$ (2.1 events/min) samples since a larger size corresponds to higher trigger efficiency. The most probable values (MPVs) of the energy deposition in 100 μm 4H-SiC layer are 25 and 55 MeV for the two different energy $\beta$ particles from the $^{90}$Sr source. There are little differences with the previous experimental result (~42 MeV) [25] due to scattering effects by the aluminum foil and LGAD, which make the ionization track slightly longer than $100\,\mu m$, but the average MPV of energy deposition from these two particles is close to the experimental result.

# TIME RESOLUTION OF 4H-SIC PIN

## Waveform Sampling

The two channels are triggered at the same time with different trigger levels for waveform sampling. Trigger$_{Ref}$ = $25\,mV$ and trigger$_{DUT}$ = $15\,mV$ are determined by noise levels (see **Figure 7**) to suppress noise spikes. **Figure 8** shows the waveforms from the LGAD (Ref) and 4H-SiC PIN (DUT), respectively. The time delay, ~5 ns, is obtained using an additional 1 m cable and is there to enhance trigger efficiency. Owing to internal gain in the LGAD, the signal of the LGAD is higher than that of 4H-SiC PIN despite the LGAD having a thinner active layer. The time resolution of the NDL LGAD is $\sigma_{Ref} = 34 \pm 1\,ps$ when the bias voltage is $U = 200\,V$ [24].

## Time Resolution

The time resolution obtained with different device sizes and reverse voltages are studied here considering the influence of capacitance and carrier velocity. For the timing method, the constant fraction discrimination (CFD) is adopted with a fraction equal to 0.5. **Figure 9** show the distribution of $\Delta T = T_{DUT} - T_{Ref}$ for different device sizes and reverse voltages. The time resolution of DUT could be extracted by $\sigma_{DUT} = \sqrt{\sigma^2_{\Delta T} - \sigma^2_{Ref}}$, where $\sigma(5\,mm \times 5\,mm,\ U = 500\,V) = 94 \pm 1\,ps$, $\sigma(5\,mm \times 5\,mm,\ U = 300\,V) = 103 \pm 1\,ps$ , and $\sigma(1.5 \times 1.5\,mm,\ U = 300\,V) = 96 \pm 2\,ps$.

At fixed size, 4H-SiC-PIN ($5 \times 5\,mm$) shows better time resolution using higher reverse voltage due to faster carrier velocity. At fixed reverse voltage, faster rising time caused by smaller capacitance improves the time resolution if the influence of the undepleted thickness in the $5 \times 5\,mm$ size device may be neglected. Meanwhile, the mean of $\Delta T$ shifts from 5.03 to 4.81 ns for different size devices due to faster rising time of devices with smaller capacitance. As it is apparent in **Figure 10**, a ~400 ps difference in the rising time leads to a ~200 ps time-shifting with 0.5 CFD fraction.

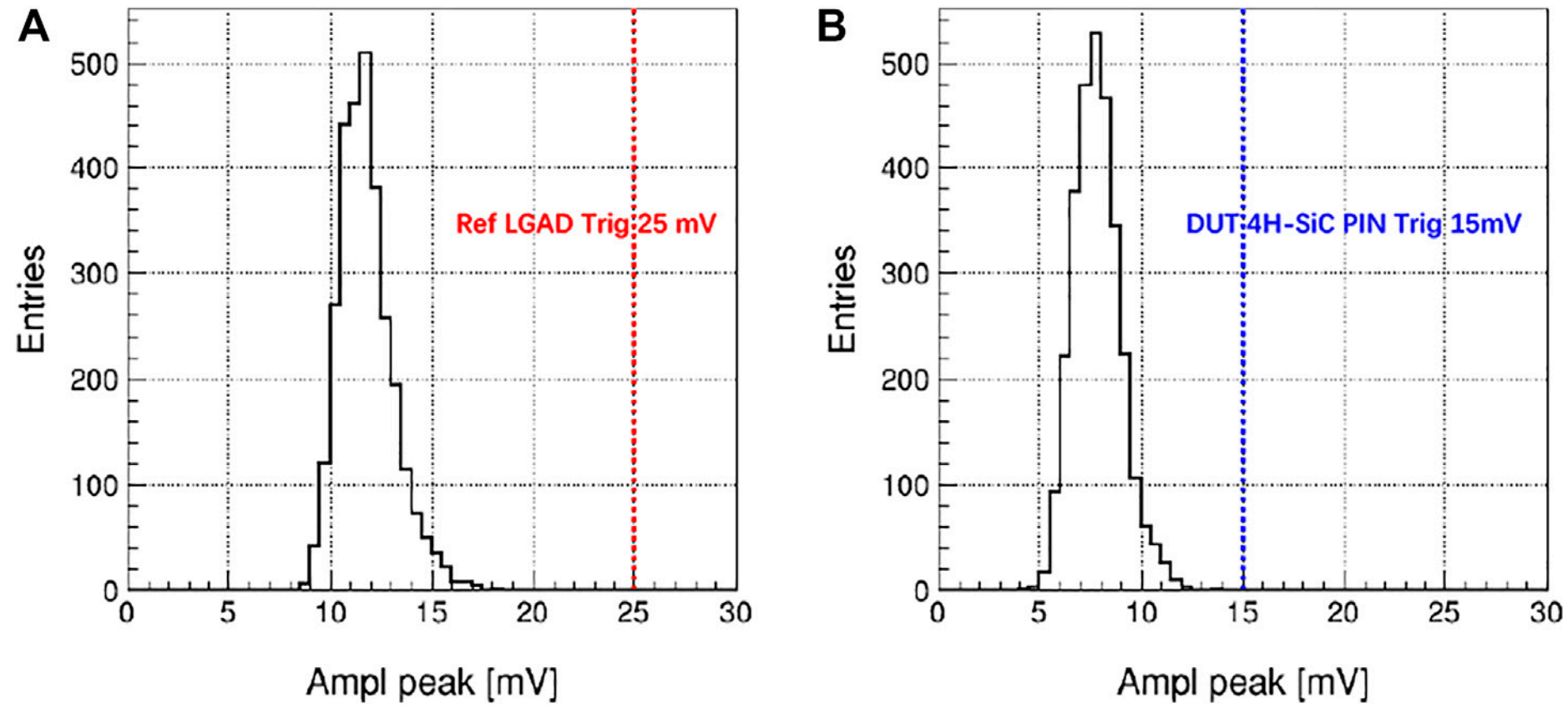

**FIGURE 7 |** Peak amplitude distribution of noise for **(A)** 33 μm Si LGAD with bias voltage U = 200 $V$ and **(B)** 100 μm 4H-SiC PIN with U = 500 $V$. The trigger levels are chosen to eliminate the fake signals from noise spikes.

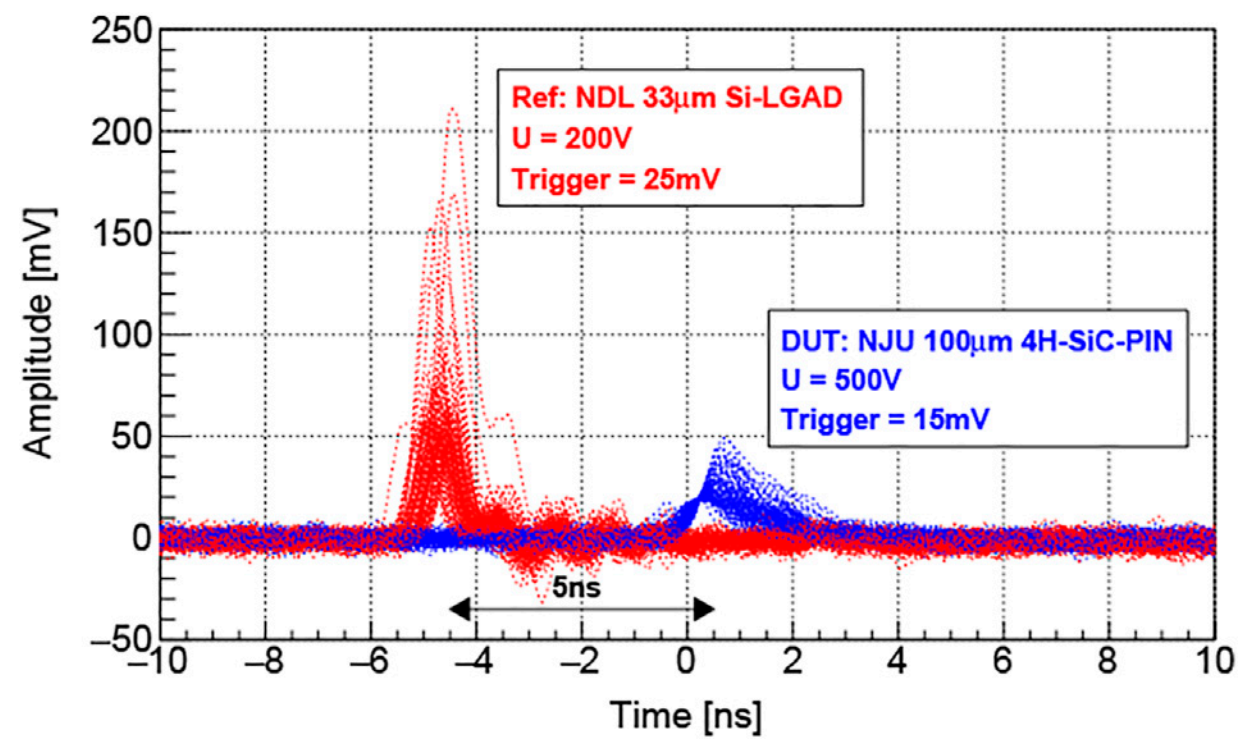

**FIGURE 8 |** Waveform sampling for Si-LGAD (red) and 4H-SiC-PIN (blue) as collected by an oscilloscope. The trigger thresholds are determined by the noise level to eliminate all noise spikes. The ~5 ns time delay of 4H-SiC-PIN signals is caused by the additional, 1 m long cable.

## SIMULATION

### Introduction of Fast Simulation Tool—RASER

We have developed a fast simulation tool, termed RASER, to study the time resolution performance of SiC detectors [21]. We use FEniCS [26], an open-source computing platform for solving partial differential equations (PDEs) to calculate the electric field and weighting potential of SiC detectors. MIPs with nonuniform charge deposition and amplitude variability are considered. The induced current is calculated by Shockley–Ramo's theorem [27] where the carrier drift is simulated using 0.1 μm steps and taking into account both magnetic field and thermal diffusion. We also assume the use of a simplified charge-sensitive amplifier (CSA) for the read-out [28].

### Weighting Field Potentials and Electric Field Calculation by FEniCS

The electric potential and weighting potential can be computed by solving Poisson's and Laplace's equations:

$$\nabla^2 \vec{U}(r) = -\frac{\rho}{\epsilon}, \tag{2}$$

$$\nabla^2 \vec{U}_w(r) = 0, \tag{3}$$

where $\vec{U}(r)$ is the electric potential, $\vec{U}_w(r)$ is the weighting potential, $\epsilon$ is the electric permittivity of SiC, and $\rho$ is the charge density. The electric and weighting field are then denoted as $\vec{E}(r) = -\nabla\vec{U}(r)$ and $\vec{E}_w(r) = -\nabla\vec{U}_w(r)$, respectively.

### Current Calculation

The induced current in the SiC detector is produced by the motion of the electron-hole pairs. The current appears when electron-hole pairs begin to move and disappears when all the pairs reach the electrode or the boundary of the detector. The instantaneous current induced by an electron or a hole can be calculated with Shockley—Ramo's theorem:

$$I = -q\vec{v}(r)\cdot\vec{E}_w(r), \tag{4}$$

where r is the position of the electron or hole, $\vec{v}(r)$ is the drift velocity, and $\vec{E}_w(r)$ is the weighting potential. The drift velocity $\vec{v}(r)$ is given by $\mu_{SiC}\cdot\vec{E}(r)$, where $\mu_{SiC}$ is the mobility in SiC. The mobility model of SiC in RASER is based on [29], and the sum of the currents induced by all electrons and holes is the total current. In the simulation, the influence of nonuniform charge deposition and impact position on the time resolution is simulated by GEANT4. The simulation contained the divergence angle from the scattering of (**Figure 4**) the upper PCB board, aluminum foil, and LGAD detector.

### Comparison Between the Simulated and Measured Time Resolution

The time resolution of the SiC detector can be expressed as follows [13]:

$$\sigma_t^2 = \sigma_{Time\ Walk}^2 + \sigma_{Landau\ Noise}^2 + \sigma_{Distortion}^2 + \sigma_{Jitter}^2 + \sigma_{TDC}^2, \tag{5}$$

where the correlations among the different items are ignored. In simulations, the time walk $\sigma_{Time\ Walk}$ is dominated by Landau variation in signal amplitude and is eliminated by the CFD method. The nonuniform charge deposition and scattering effects cause Landau noise $\sigma_{Landau\ Noise}$. The nonuniform weighting potential and the electric field cause signal distortion $\sigma_{Distortion}$. The electronics noise, which leads to $\sigma_{Jitter}$ and $\sigma_{TDC}$, is dominated by the binning of signal waveforms. All these time resolution contributions, except the distortion term, are considered in the simulation process.

We simulated the time resolution of the NJU detector with $5\,mm \times 5\,mm$ sizes with a 500 V bias voltage at room temperature T = 300 $K$. In these conditions, the average carrier velocities are $V_{electron} = 150\,\mu m/ns$ and $V_{hole} = 50\,\mu m/ns$. We use RASER to model an ideal planar detector and calculate the electric field so that $\sigma_{Distortion}$ is not considered in the simulation. Based on GEANT4 simulation, the nonuniform charge deposition and scattering effects are applied in RASER to reproduce the contribution of $\sigma_{Landau\ Noise}$. The random noise is added in each signal waveform to estimate $\sigma_{Jitter}$. The contribution of $\sigma_{TDC}$ has been considered in binning the signal (the same bin interval 50 $ps/bin$ with sampling time step). The program has been validated by comparing the induced current for MIPs of RASER simulations and measured signals. **Figure 11** shows the comparison between waveforms obtained from RASER and TCAD simulation and measured data, where the mobility model applied in TCAD is the Masetti model with parameters from [30, 31]. Good consistency is found in all three cases.

The results of simulations for the time resolution of the NJU detector are shown in **Figure 12**. We use 20,000 events, and the same CFD fraction 0.5 as in the measurements is used to obtain the time of arrival (ToA). The time resolution obtained by Gaussian fitting is (73 ± 1 *ps*), where the simulated $\sigma_{Jitter}$ is 66 *ps* and leftover $\sigma_{Landau\ Noise}$ and $\sigma_{TDC}$ is 30 *ps* in total due to $\sigma_{Time\ Walk}$ being absent. The simulated time resolution is (24 ± 1) *ps* less than the measurement result. The origins of this difference are likely because there is no shielding box of the beta source measurement, which leads to an increase in time resolution. RASER software can effectively simulate the time resolution of SiC to some extent, but measurements and simulations both need further optimization.

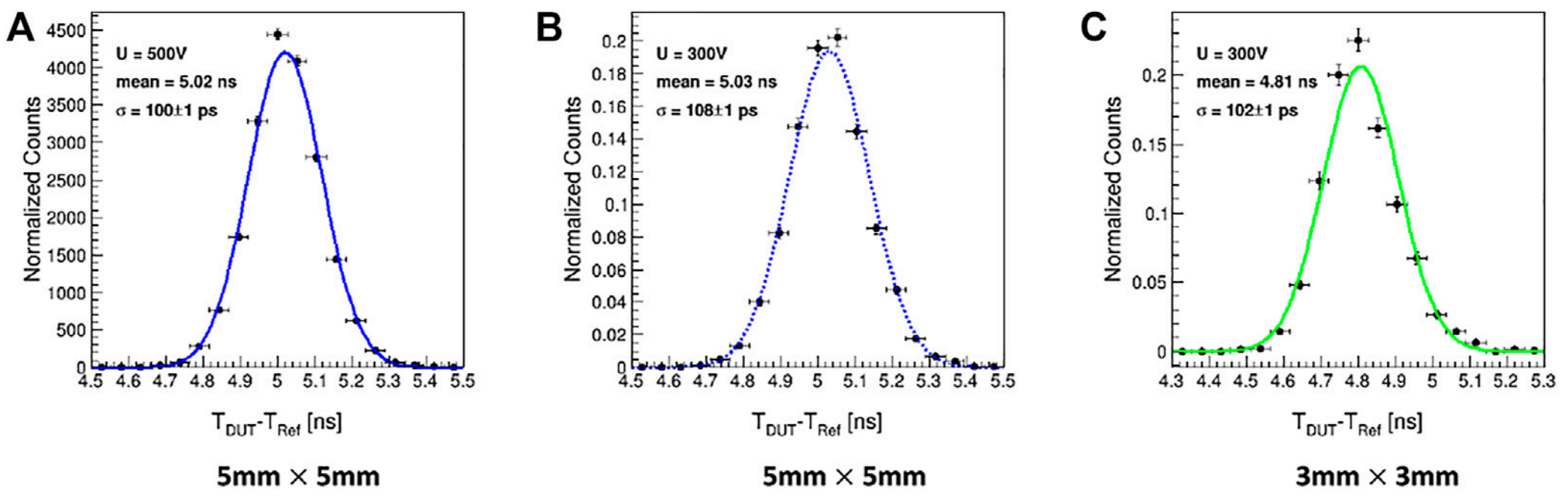

**FIGURE 9 |** $\Delta T = T_{DUT} - T_{Ref}$ distributions after applying the CFD method with fraction = 0.5. **(A)**: 5 $mm$ × 5 $mm$ and U = 500 $V$, **(B)**: 5 $mm$ × 5 $mm$ and U = 300 $V$, and **(C)**: 1.5 $mm$ × 1.5 $mm$ and U = 300 $V$. The $\sigma_{\Delta T}$ is extracted by Gaussian fitting.

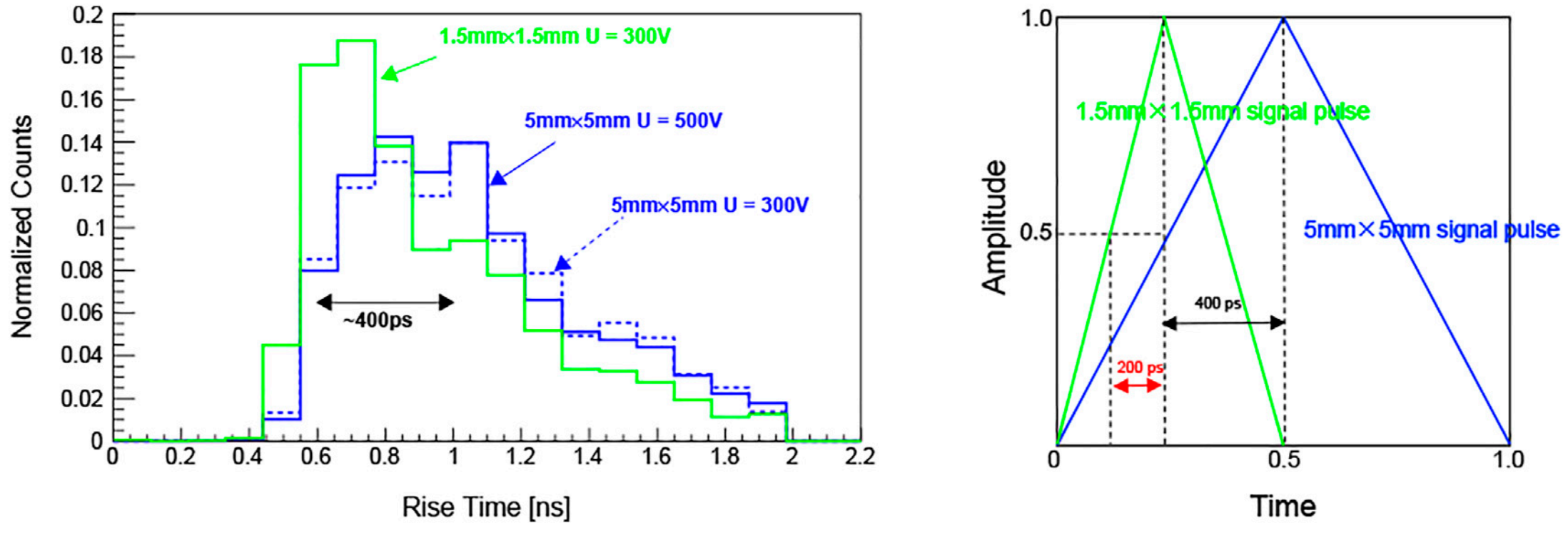

**FIGURE 10 |** Qualitative explanation of the origin of the shifts of the mean of $\Delta T$. The left figure shows the distribution of signal rising time for different size devices. The right figure indicates that a faster rising time leads to a smaller ToT (time over threshold), so the mean of $\Delta T$ is shifted.

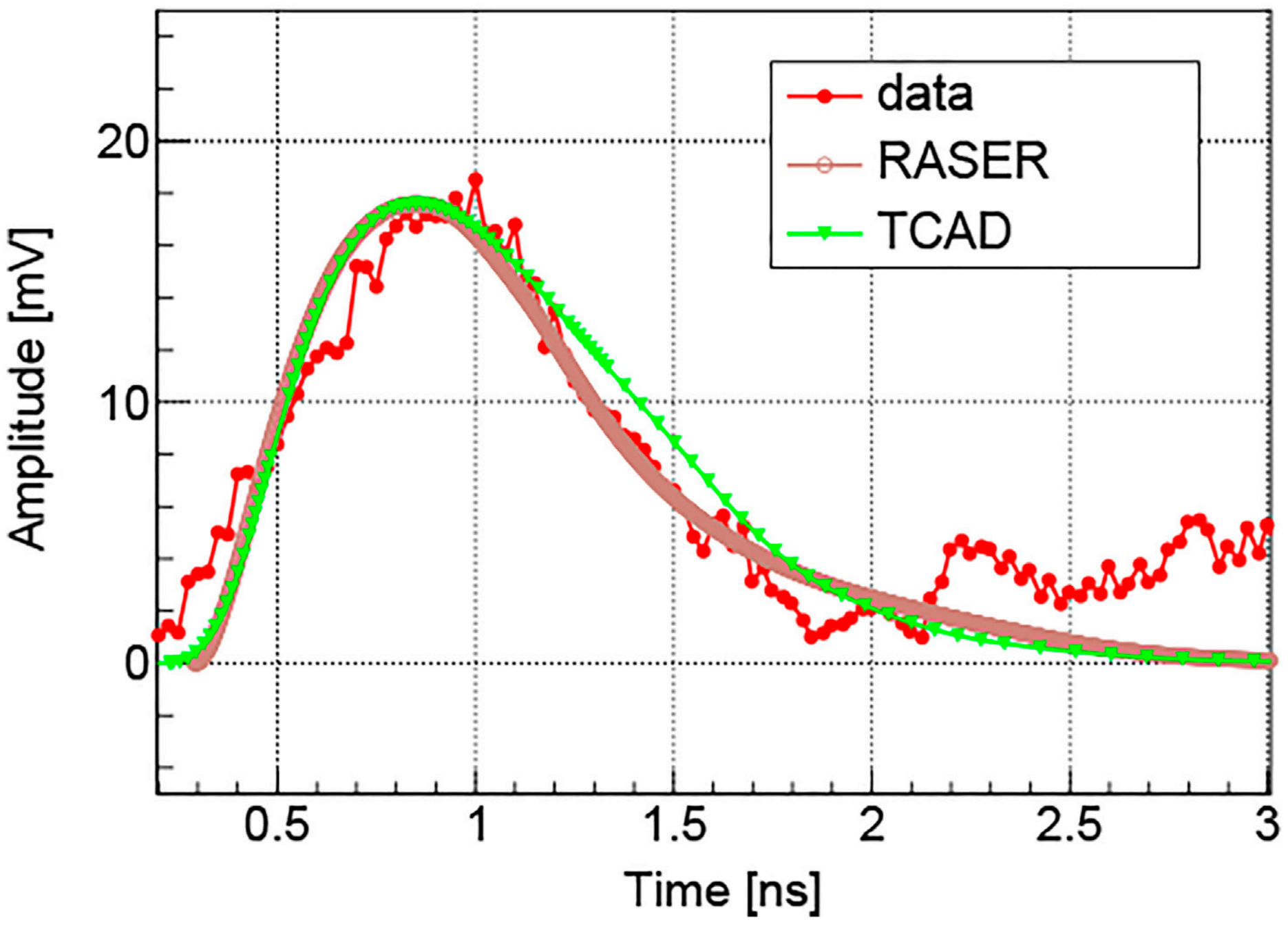

**FIGURE 11 |** (Color online) Waveform comparison: measured data (brown) and RASER simulation (green).

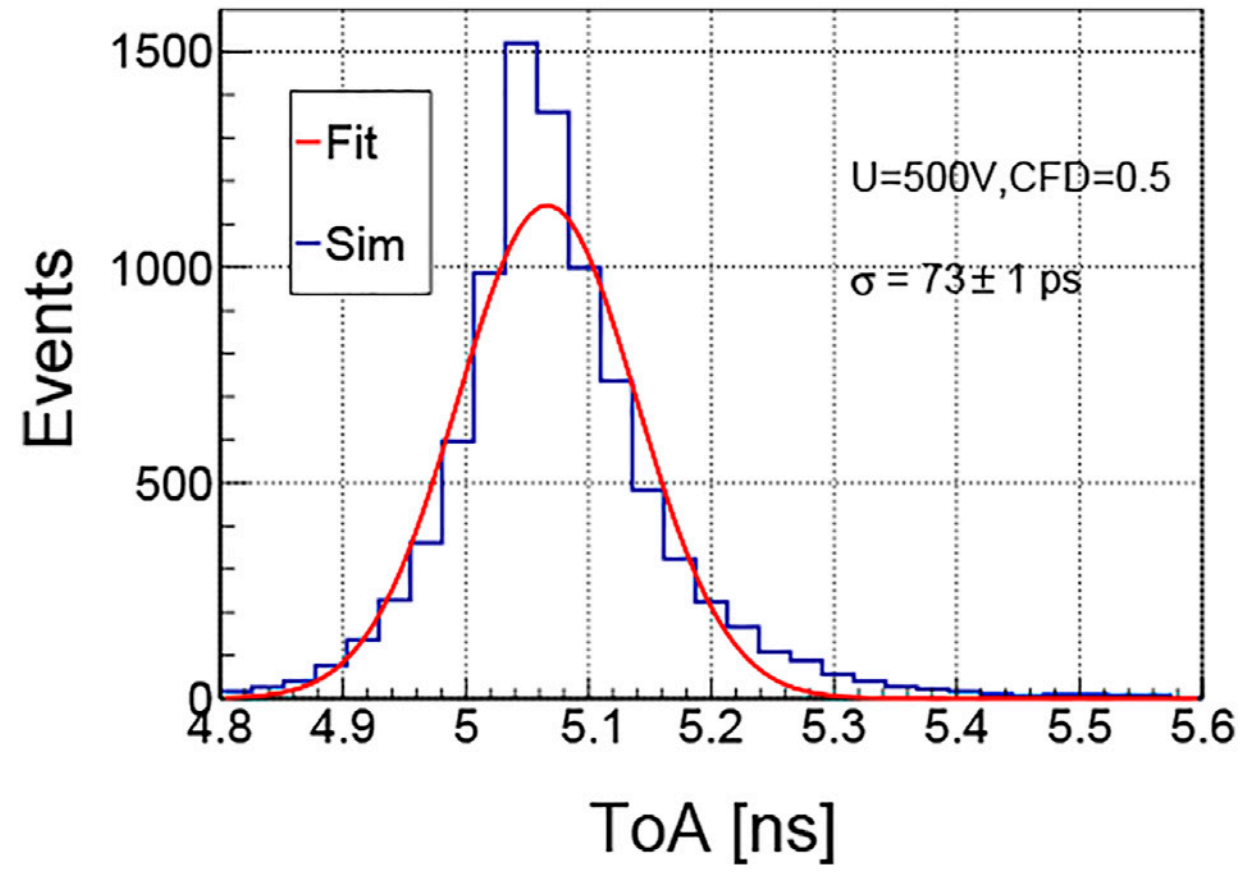

**FIGURE 12 |** (Color online) Time resolution simulation results of the 5 mm×5 mm size NJU detector.

## CONCLUSION

The best time resolution of the NJU 5 mm×5 mm 4H-SiC-PIN detector with the $^{90}$Sr source is 94 ± 1 *ps*. On using higher reverse voltage and smaller capacitance, better time resolution is obtained. The waveform simulated by RASER has been validated against measurements. The simulated time resolution indicates that all the leading contributions of the test system should be considered to obtain reliable results. Our study about measured and simulated time resolutions is useful to develop ultrafast 4H-SiC LGAD and 3D 4H-SiC detectors [32] which are expected to achieve improved time resolution in the near future.

## DATA AVAILABILITY STATEMENT

The raw data supporting the conclusion of this article will be made available by the authors, without undue reservation.

## AUTHOR CONTRIBUTIONS

Conceptualization: XS. Experimental setup and measurement: TY, JZ, and SX. Software development: TY, YT, SX, KL, RK, and JL. TCAD Simulation: TY and BW. Developing the SiC sample: HL and QL. Writing and original draft preparation: TY and YT. Supervision: MZ, XZ, and CW. All authors have read and agreed to the published version of the manuscript.

## FUNDING

This research is supported by the National Natural Science Foundation of China (No.11961141014); and the Program of Science and Technology Development Plan of Jilin Province of China under Contract No. 20210508047RQ.

## ACKNOWLEDGMENTS

This work is carried out in the CERN RD50 framework. We would like to acknowledge the TeledyneLecroy Beijing office for providing us with a fast sampling oscilloscope, Gregor Kramberger for helpful discussions about the simulation method in KDetSim, and the IHEP HGTD group for the hardware support and for the suggestions concerning time resolution measurements.

## SUPPLEMENTARY MATERIAL

The Supplementary Material for this article can be found online at: https://www.frontiersin.org/articles/10.3389/fphy.2022.718071/full#supplementary-material

**Conflict of Interest:** The authors declare that the research was conducted in the absence of any commercial or financial relationships that could be construed as a potential conflict of interest.

**Publisher's Note:** All claims expressed in this article are solely those of the authors and do not necessarily represent those of their affiliated organizations, or those of the publisher, the editors, and the reviewers. Any product that may be evaluated in this article, or claim that may be made by its manufacturer, is not guaranteed or endorsed by the publisher.